\documentclass[%
reprint,
superscriptaddress,
amsmath,amssymb,
aps,
prb,
floatfix,
]{revtex4-1}

\usepackage{graphicx}
\usepackage{dcolumn}
\usepackage{bm}
\usepackage{hyperref}
\usepackage{color}
\usepackage{multirow}
\usepackage{rotating}

\begin{document}

\title{Towards predictive band gaps for halide perovskites: Lessons from one-shot and eigenvalue self-consistent GW}

\author{Linn Leppert}
\affiliation{
Department of Physics, University of Bayreuth, 95440 Bayreuth, Germany
}
\author{Tonatiuh Rangel}
\affiliation{
Department of Physics, University of California Berkeley, Berkeley, CA 94720, United States
}
\affiliation{Molecular Foundry, Lawrence Berkeley National Laboratory, Berkeley, CA 94720, United States}
\author{Jeffrey B. Neaton}
\affiliation{
Department of Physics, University of California Berkeley, Berkeley, CA 94720, United States
}
\affiliation{Molecular Foundry, Lawrence Berkeley National Laboratory, Berkeley, CA 94720, United States}
\affiliation{Kavli Energy NanoSciences Institute at Berkeley, Berkeley, CA 94720, United States}
\date{\today}

\begin{abstract}
Halide perovskites constitute a chemically-diverse class of crystals with great promise as photovoltaic absorber materials, featuring band gaps between about 1 and 3.5\,eV depending on composition. Their diversity calls for a general computational approach to predicting their band gaps. However, such an approach is still lacking. Here, we use density functional theory (DFT) and \textit{ab initio }many-body perturbation theory within the GW approximation to compute the quasiparticle or fundamental band gap of a set of ten representative halide perovskites: CH$_3$NH$_3$PbI$_3$ (MAPbI$_3$), MAPbBr$_3$, CsSnBr$_3$, (MA)$_2$BiTlBr$_6$, Cs$_2$TlAgBr$_6$, Cs$_2$TlAgCl$_6$, Cs$_2$BiAgBr$_6$, Cs$_2$InAgCl$_6$, Cs$_2$SnBr$_6$, and Cs$_2$Au$_2$I$_6$. Comparing with recent measurements, we find that a standard generalized gradient exchange-correlation functional can significantly underestimate the experimental band gaps of these perovskites, particularly in cases with strong spin-orbit coupling (SOC) and highly dispersive band edges, to a degree that varies with composition. We show that these nonsystematic errors are inherited by one-shot G$_0$W$_0$ and eigenvalue self-consistent GW$_0$ calculations, demonstrating that semilocal DFT starting points are insufficient for MAPbI$_3$, MAPbBr$_3$, CsSnBr$_3$, (MA)$_2$BiTlBr$_6$, Cs$_2$TlAgBr$_6$, and Cs$_2$TlAgCl$_6$. On the other hand, we find that DFT with hybrid functionals leads to an improved starting point and GW$_0$ results in better agreement with experiment for these perovskites. Our results suggest that GW$_0$ with hybrid functional-based starting points are promising for predicting band gaps of systems with large SOC and dispersive bands in this technologically important class of semiconducting crystals.
\end{abstract}

\maketitle
\section{Introduction}
Solar cells based on hybrid halide perovskites have become promising contenders in the race for maximum power conversion efficiency (PCE). Easy to process and possessing wide chemical tunability \cite{Stranks2015b,Petrus2017}, single cell devices with halide perovskite absorbers exhibit remarkable PCEs of more than 23\% \cite{nrel2017}. Although the origin of their high efficiencies is not well understood, the electronic structure of materials like methylammonium lead iodide, CH$_3$NH$_3$PbI$_3$ (MAPbI$_3$), unequivocally plays a central role in their success. MAPbI$_3$ has a direct room temperature band gap of 1.6\,eV \cite{Kojima2009}, only slightly larger than the ideal Shockley-Queisser band gap of $\sim$1.3\,eV. Additionally, its valence band maximum (VBM) and conduction band minimum (CBM) are highly dispersive with similar and relatively low effective masses, suggesting facile electron and hole transport \cite{Xing2013,Stranks2013a,Miyata2015a}.

Nonetheless, the presence of toxic Pb, the relative scarcity of I, and stability issues have led to an ongoing quest for more stable, environmentally benign, and earth-abundant materials with similar electronic properties \cite{Brandt2015,Savory2016}. Two strategies for finding alternatives to MAPbI$_3$ are a) to keep the simple perovskite stoichiometry ABX$_3$ and substitute the central metal ion Pb$^{+2}$ with other ions with nominal oxidation state +2 (Ref.~\citenum{Filip2015a}), and b) to explore double perovskites \cite{Faber2016}, a generalization of ABX$_3$ typically crystallizing in structures with Fm$\bar{3}$m symmetry with two different B site ions. Among double perovskites with stoichiometry A$_2$BB'X$_6$, Cs$_2$BiAgBr$_6$ has been synthesized and shown to have an indirect band gap between 1.8\,eV\cite{Slavney2016a} and 2.2\,eV \cite{McClure2016}; (MA)$_2$BiTlBr$_6$ \cite{Slavney2017}, Cs$_2$InAgCl$_6$ \cite{Volonakis2017}, and Cs$_2$TlAgX$_6$ (X=Br, Cl) \cite{Slavney2018a} have all also been synthesized and reported to have direct band gaps ranging from 1.0\,eV to 3.2-3.5\,eV. Furthermore, the double perovskite stoichiometry allows the realization of vacancy-ordered perovskites like Cs$_2$Sn(IV)Br$_6$, and charge-ordered perovskites like Cs$_2$Au(III)Au(I)I$_6$, both with direct band gaps of 2.7-2.9\,eV \cite{Kaltzoglou2016,Lee2017a}, and 1.3\,eV\cite{Debbichi2018}, respectively.

First principles calculations have made important contributions to the understanding of halide perovskites, including the prediction of new materials that were subsequently synthesized \cite{Volonakis2017}. While density functional theory (DFT) is a standard tool to calculate ground state properties with good accuracy, its common approximations underestimate quasiparticle (QP), or fundamental, band gaps of solids \cite{Perdew2017}, often to different degrees, depending on the material. Accurate QP energies, and thus fundamental band gaps, can be obtained in principle using Green's function-based \textit{ab initio} many-body perturbation theory (MBPT). In MBPT, the QP eigensystem is the solution of one-particle equations solved in the presence of a non-local, energy-dependent self-energy operator $\Sigma$ \cite{Fetter1971}. In practice, $\Sigma$ is most commonly approximated as $iGW$, the first-order term in an expansion of $\Sigma$ in the screened Coulomb interaction $W$ \citep{Aryasetiawan1998,Hedin1999}, where $G$ is the one-particle Green's function. And instead of evaluating $\Sigma$ self-consistently, the QP eigenvalues are frequently calculated in a computationally less demanding one-shot approach referred to as G$_0$W$_0$, or using partial or full eigenvalue self-consistency, i.e., GW$_0$ or GW. In G$_0$W$_0$, $G_0$ and $W_0$ are constructed from the DFT generalized Kohn-Sham (gKS) eigensystem, and the gKS eigenvalues are perturbatively corrected. In eigenvalue self-consistent GW, the gKS eigenvalues used to construct G and/or W are replaced with those from the output of a prior GW step; the self-energy corrections are then iterated until the QP eigenvalues converge.

One-shot G$_0$W$_0$ has been successful in significantly improving band gap predictions for many systems, from molecules to solids \citep{Hybertsen1985,Blase1995,Shih2010,Klimes2014}. However, as G$_0$W$_0$ is a perturbative method, it can be sensitive to the quality of the underlying gKS "starting point", and therefore the exchange-correlation functional used to construct it. This sensitivity can lead to different band gaps for different starting points, and indeed, this has been shown to be the case in prior G$_0$W$_0$ calculations of band gaps of halide perovskites. For MAPbI$_3$ a number of GW calculations have been reported, although the different technical implementations and lattice parameters used in these calculations complicate their comparison. G$_0$W$_0$ calculations with a PBE starting point for MAPbI$_3$ and MAPbBr$_3$, in which spin-orbit coupling (SOC) was included in the calculation of $G$, but $W$ was calculated scalar-relativistically, resulted in very good agreement with experiment \cite{Umari2014,Mosconi2016a}. Brivio et al.~showed, with an LDA starting point and including SOC self-consistently in the calculation of $G$ and $W$, that QP selfconsistency in both eigenvalues and wavefunctions is necessary to obtain reliable band gaps for MAPbI$_3$ \cite{Brivio2014}. Filip et al.~used a scissor-correction from an LDA starting point to achieve self-consistency in the QP eigenvalues and good agreement with experiment \cite{Filip2014d}. Wiktor et al.~discussed the effect of QP self-consistency, vertex corrections, thermal vibrations, and disorder for the simple inorganic perovskites CsPbX$_3$ and CsSnX$_3$ (X=I, Br, Cl) \cite{Wiktor2017a}, concluding that all these effects need to be accounted for to obtain good agreement with experiment. While QP self-consistent GW calculations are a promising strategy for obtaining reliable band gaps for some halide perovskites\cite{Huang2013,Brivio2014,Wiktor2017a}, their computational expense is greater than that of one-shot GW, and there is quantitative disagreement between gaps reported from different studies using the approach. For double perovskites there is, to the best of our knowledge, only one GW study and its G$_0$W$_0$@LDA band gap was computed to be in good agreement with experiment \cite{Filip2016}. Concluding from this significant body of prior work, a one-size-fits-all solution for predicting band gaps and band gap trends has yet to be demonstrated for this growing and chemically diverse class of compounds, motivating the need for further GW studies on a broader class of perovskite materials.

Here, we assess the predictive power of the one-shot G$_0$W$_0$ and eigenvalue self-consistent GW$_0$ approaches for band gaps of a series of halide perovskites, reporting G$_0$W$_0$ and GW$_0$ gaps for perovskites of different stoichiometry and chemical compositions with different DFT starting points. We find that for halide perovskites like MAPbX$_3$ (X=I, Br), CsSnBr$_3$, (MA)$_2$BiTlBr$_6$, and Cs$_2$TlAgX$_6$ (X=Br, Cl) with strong spin-orbit interactions and/or dispersive band edges, eigensystems computed with semilocal exchange-correlation functionals such as PBE are insufficient starting points for G$_0$W$_0$ and GW$_0$, resulting in predicted band gaps greatly underestimated relative to experiment. We demonstrate that significantly improved agreement with experiment and improved band gap trends can be reached for such systems using hybrid functional-based starting points, and our calculations should be useful for future one-shot and self-consistent GW studies of halide perovskites.

In the following, we compute the band gaps of the simple perovskites MAPbX$_3$ (X=I, Br) and CsSnBr$_3$ in their high-temperature phase with Pm$\bar{3}$m symmetry and experimental lattice parameters. We also compute gaps of the double perovskites Cs$_2$BiAgBr$_6$, (MA)$_2$BiTlBr$_6$, Cs$_2$TlAgX$_6$ (X=Br, Cl), Cs$_2$InAgCl$_6$, and Cs$_2$SnBr$_6$ which all crystallize in Fm$\bar{3}$m symmetry, and Cs$_2$Au$_2$I$_6$ with I4/mmm symmetry and a regular rocksalt ordering of the B and B' cations \cite{Faber2016}. Fig.~\ref{fig:unitcell}a and \ref{fig:unitcell}b show the primitive unit cells of MAPbBr$_3$, and (MA)$_2$BiTlBr$_6$ as examples. All compounds studied here have been synthesized and characterized experimentally. They represent the diversity of chemical compositions and electronic structures that is characteristic of halide perovskites, with measured band gaps ranging from 0.95\,eV for Cs$_2$TlAgBr$_6$ to 3.30-3.53\,eV for Cs$_2$InAgCl$_6$. We note that all experimental band gaps cited here are optical gaps. MAPbX$_3$ (X=I, Br), CsSnBr$_3$, and Cs$_2$TlAgX$_6$ (X=Br, Cl) possess exciton binding energies $<$50\,meV \citep{Miyata2015a, Slavney2018a}, while the exciton binding energy of Cs$_2$AgBiBr$_6$ has recently been reported to be $\sim$270\,meV \citep{Kentsch2018}.

\section{Methodology}
\subsection{Computational Details}
\begin{figure*}[t]
 \includegraphics[width=13cm]{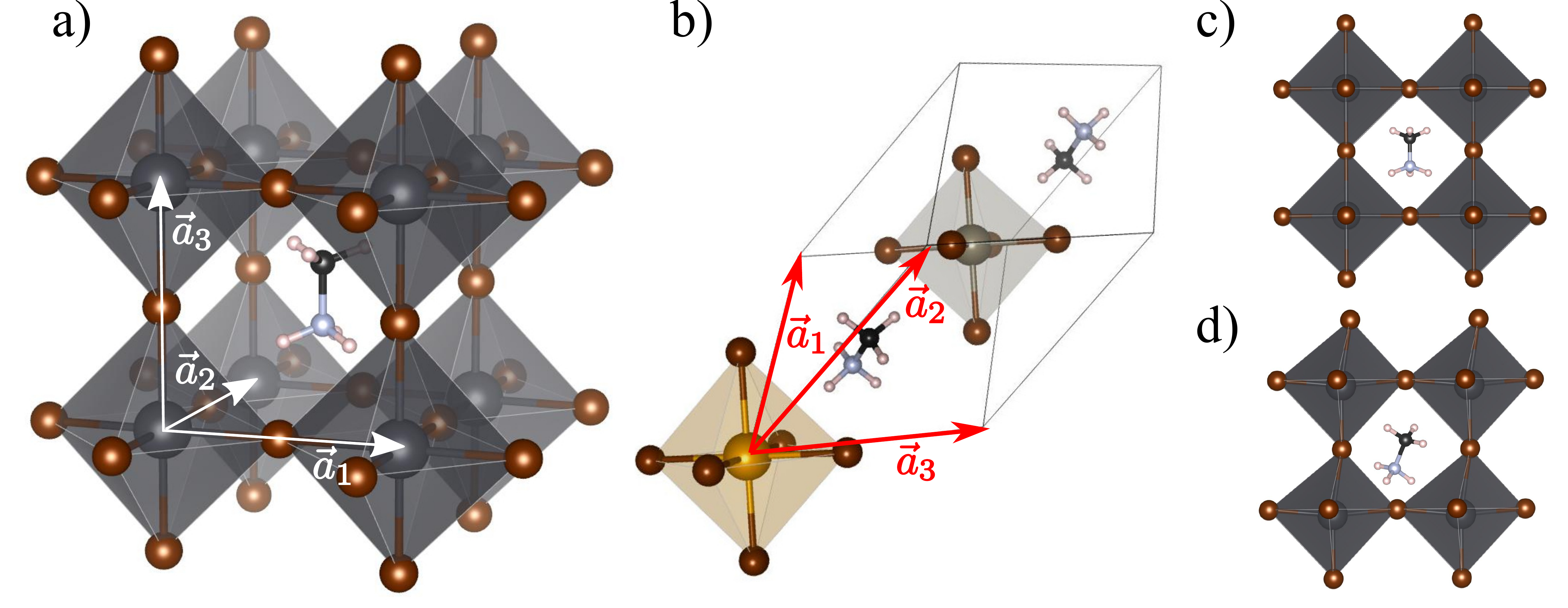}
 \caption{a) Unit cell of MAPbBr$_3$ with Pm$\bar{3}$m symmetry. b) Primitive unit cell of (MA)$_2$BiTlBr$_6$ with Bi in orange, Tl in gray, Br in brown, C in black, N in blue and H in white. c) Experimental, and d) geometry-optimized \cite{Bokdam2016} structure of MAPbBr$_3$ demonstrating relatively large deviations from the experimentally demonstrated Pm$\bar{3}$m symmetry. Since we replaced MA by Cs in all calculations, the orientation of the molecule does not affect calculated band gaps.}
  \label{fig:unitcell}
\end{figure*}
In all of our calculations, we use experimental lattice parameters and space groups (Table \ref{tab:lattice}) and replace the MA cations by Cs (CIFs of all structures can be found in the Supplemental Material): Macroscopic alignment of the MA molecules breaks centrosymmetry for MA-based perovskites (Section \ref{subsec:structure})\cite{Leppert2016}. Together with strong SOC this inversion symmetry breaking leads to an effective magnetic field driving a spin splitting \cite{Kepenekian2017,Rashba1960,Dresselhaus1955}. However, at room and higher temperature and without applied fields the molecules are believed to rotate quasi-freely and MAPbX$_3$ (X=I, Br) and (MA)$_2$BiTlBr$_6$ to be centrosymmetric \cite{Frohna2018a}. For (MA)$_2$BiTlBr$_6$ we find that replacing the two MA molecules (oriented as shown in Fig.~\ref{fig:unitcell}b) by Cs changes the calculated band gap by less than 0.1\,eV both at the DFT and GW level. This simplification also allows us to use a primitive unit cell with one formula unit because the Cs cation effectively mimics the on-average centrosymmetric structure of these compounds at room and higher temperatures. We note, however, that we neglect the effect of thermal or dynamical fluctuations of the atomic structure on the band gap, as further discussed in Section \ref{subsec:structure}.
\begin{table}
\caption{Space groups and experimental lattice parameters of the primitive unit cell of all systems considered in this work.}
\begin{center}
\begin{tabular}{ccc}
\hline \hline
System & Space Group & Lattice Parameters (\AA) \\
\hline
MAPbI$_3$ & $Pm\bar{3}m$ & a=6.33\cite{Poglitsch1987} \\
MAPbBr$_3$ & $Pm\bar{3}m$ & a=5.93\cite{Noh2013} \\
CsSnBr$_3$ & $Pm\bar{3}m$ & a=5.80\cite{Barrett1971}\\ \hline
(MA)$_2$BiTlBr$_6$ & $Fm\bar{3}m$ & a=11.92\cite{Slavney2017}\\
Cs$_2$TlAgBr$_6$ & $Fm\bar{3}m$ & a=11.10\cite{Slavney2018a}\\
Cs$_2$TlAgCl$_6$ & $Fm\bar{3}m$ & a=10.56\cite{Slavney2018a}\\
Cs$_2$BiAgBr$_6$ & $Fm\bar{3}m$ & a=11.25\cite{Slavney2016a} \\
Cs$_2$InAgBr$_6$ & $Fm\bar{3}m$ & a=10.47\cite{Volonakis2017}\\
Cs$_2$SnBr$_6$ & $Fm\bar{3}m$ & a=10.77\cite{Ketelaar1937}\\
Cs$_2$Au$_2$I$_6$ & $I4/mmm$ & a=b=8.28, c=12.09\cite{Debbichi2018} \\
\hline \hline
\end{tabular}
\end{center}
\label{tab:lattice}
\end{table}

We start by performing DFT calculations using the PBE functional as implemented in \textsc{vasp} \cite{Kresse1994,Kresse1996} for all ten compounds considered here. We indicate the gKS eigenvalues as $E^{\text{gKS}}_{nk}$ to distinguish them from the QP eigenvalues from our GW calculations. SOC is included self-consistently for all compounds apart from Cs$_2$TlAgX$_6$ (X=Br, Cl) and Cs$_2$InAgCl$_6$; for these three compounds, for which we do not use SOC, self-consistent SOC changes the band gap by less than 0.05\,eV. We use projector augmented wave (PAW) potentials as described in Ref.~\citenum{Klimes2014}, including semicore electrons explicitly in our calculations for all elements apart from Cs and Cl (see Section \ref{subsec:pseudo}). All Brillouin zone integrations for our DFT calculations are performed using $\Gamma$-centered 2$\times$2$\times$2 k-point grids with a cutoff energy of 500\,eV for the plane wave expansion of the wavefunctions. 

Our QP eigenvalues are obtained via approximate solution of the Dyson equation
\begin{equation}
\left[-\frac{1}{2}\nabla^2 + V_{\text{ion}} + V_{\text{H}} + \Sigma(E_{nk}^{\text{QP}}) \right]\psi_{nk}^{\text{QP}} = E_{nk}^{\text{QP}}\psi_{nk}^{\text{QP}}.
\end{equation}
Here, $V_{\text{ion}}$ is the ionic (pseudo)potential, $V_{\text{H}}$ is the Hartree potential and $\Sigma=iGW$ in the GW approximation. $E_{nk}^{\text{QP}}$ and $\psi_{nk}^{\text{QP}}$ are QP energies and wavefunctions, respectively. In the G$_0$W$_0$ approach, QP corrections are calculated to first order in $\Sigma$ as
\begin{equation}
\label{equation1}
E_{nk}^{\text{QP}} = E_{nk}^{\text{gKS}} + \langle \psi_{nk} | \Sigma(E_{nk}^{\text{QP}}) - V_{xc} | \psi_{nk} \rangle,
\end{equation}
where $V_{\text{xc}}$ is the exchange-correlation potential. Here, $\Sigma$ is computed using gKS eigenvalues and eigenfunctions to construct $G_0$ and $W_0$, and evaluated at the QP energy $E^{\text{QP}}_{nk}$. Note that our one-shot calculations assume $\psi_{nk} \approx \psi_{nk}^{\text{QP}}$. We use the notation G$_0$W$_0$@gKS to refer to G$_0$W$_0$ based on the gKS eigensystem computed with the exchange-correlation functional $E_{\text{xc}}^{\text{gKS}}$. When used, SOC is included in the construction of both G$_0$ and W$_0$. We use the \textsc{vasp} code for all full-frequency GW calculations. Our extensive convergence tests are discussed in Section \ref{subsec:convergence}.

\subsection{Effect of pseudopotential}
\label{subsec:pseudo}
The effect of using pseudopotentials including semicore electrons on GW calculations of QP energy levels is well documented in the literature \citep{Rohlfing1995,Tiago2004,Gomez-Abal2008,Klimes2014}. Here we tested two different valence electron configurations PAW1 and PAW2 (Table \ref{tab:pseudos}) for the three halide perovskites MAPbBr$_3$, (MA)$_2$BiTlBr$_6$, and Cs$_2$BiAgBr$_6$. PAW2 includes semicore states for all elements apart from Cs; Cs orbitals do not contribute to the electronic states close to the band edges. Furthermore, we compare PAW1 and PAW2 PBE band gaps with results obtained with the all-electron full-potential linearized augmented-plane wave (FP-LAPW) code Elk v4.3.06 \cite{elk2017}. In Elk we use the highly converged \textit{highq} setting and 2$\times$2$\times$2 k-point grids to obtain a similar level of convergence as in \textsc{vasp}.
\begin{table}
\caption{Valence configurations of PAW potentials PAW1 and PAW2 used in \textsc{vasp}.}
\begin{center}
\begin{tabular}{ccc}
\hline \hline
element & PAW1 & PAW2 \\ \hline
Pb & 5d$^{10}$6s$^2$6p$^2$ & 5s$^2$5p$^6$5d$^{10}$6s$^2$6p$^2$ \\
Bi & 6s$^2$6p$^3$ & 5s$^2$5p$^6$5d$^{10}$6s$^2$6p$^3$ \\
Tl & 6s$^2$6p$^1$ & 5s$^2$5p$^6$5d$^{10}$6s$^2$6p$^1$ \\
Ag & 4s$^2$4p$^6$4d$^{10}$5s$^1$, & 4s$^2$4p$^6$4d$^{10}$5s$^1$ \\
Cs & 5s$^2$5p$^6$6s$^1$ & 5s$^2$5p$^6$6s$^1$ \\
Br & 4s$^2$4p$^5$ & 3s$^2$3p$^6$3d$^{10}$4s$^2$4p$^5$ \\ \hline \hline
\end{tabular}
\end{center}
\label{tab:pseudos}
\end{table}

We find that when SOC is included self-consistently, the presence of semicore electrons has a large effect on PBE band gaps, in particular for MAPbBr$_3$ and (MA)$_2$BiTlBr$_6$, for which the difference between the PAW1 and the FP-LAPW band gap is $\sim$0.2\,eV. The effect is less pronounced for Cs$_2$BiAgBr$_6$, with only 0.05\,eV difference between PAW1 and FP-LAPW. PAW2 consistently results in PBE band gaps closer to the all-electron result, although for (MA)$_2$BiTlBr$_6$ the PAW2 band gap is still 0.1\,eV lower than the FP-LAPW band gap. When SOC effects are neglected, we find that band gap differences between PAW1, PAW2 and FP-LAPW are below 0.02\,eV for all three systems. This finding suggests that for halide perovskites with strong SOC, the inclusion of semicore electrons in the pseudopotential can significantly affect the DFT gKS eigenvalues and band gap.

\begin{table}
\caption{PBE band gaps (in eV) calculated using PAW1 and PAW2 with \textsc{vasp}, and FP-LAPW with Elk, both with and without including SOC self-consistently.}
  \begin{tabular}{ccccccc}
    \hline \hline
    \multirow{2}{*}{System} &
      \multicolumn{3}{c}{with SOC} &
      \multicolumn{3}{c}{w/o SOC} \\
    & PAW1 & PAW2 & Elk & PAW1 & PAW2 & Elk \\
    \hline
    MAPbBr$_3$ & 0.39 & 0.55 & 0.59 & 1.58 & 1.58 & 1.58 \\
    (MA)$_2$BiTlBr$_6$ & 0.50 & 0.60 & 0.70 & 1.62 & 1.62 & 1.61 \\
    Cs$_2$BiAgBr$_6$ & 1.06 & 1.09 & 1.11 & 1.24 & 1.24 & 1.22 \\
    \hline \hline
  \end{tabular}
  \label{tab:meanfieldgaps}
\end{table}
\begin{table}
\caption{G$_0$W$_0$@PBE band gaps with PAW1 and PAW2 (in eV) calculated with \textsc{vasp}, including SOC.}
\begin{center}
\begin{tabular}{ccc}
\hline \hline
System & PAW1 & PAW2 \\
\hline
MAPbBr$_3$ & 1.03 & 1.49 \\
(MA)$_2$BiTlBr$_6$ & 1.24 & 1.39 \\
Cs$_2$BiAgBr$_6$ & 1.87 & 2.01 \\
\hline \hline
\end{tabular}
\end{center}
\label{tab:gwgaps}
\end{table}

Table \ref{tab:gwgaps} compares G$_0$W$_0$@PBE band gaps calculated with PAW1 and PAW2 including SOC. Our findings are consistent with previous reports \cite{Filip2014d,Scherpelz2016,Wiktor2017a}: PAW potentials with valence configurations including all semicore electrons yield significantly larger G$_0$W$_0$@PBE band gaps -- up to $\sim$0.5\,eV for MAPbBr$_3$. In what follows, we use PAW2 potentials for all calculations.

\subsection{Convergence of GW calculations}
\label{subsec:convergence}
GW calculations feature several interdependent convergence parameters, such as the number of unoccupied states involved in the calculation of the irreducible polarizability and the self-energy, and the plane wave energy cutoff for the dielectric matrix $\epsilon_{\bm G, \bm G'}(\bm q)$. In what follows, we describe our tests of the convergence of the G$_0$W$_0$@PBE band gap as a function of these parameters for MAPbI$_3$ and (MA)$_2$BiTlBr$_6$. Because of the computational demands associated with hybrid functional starting points, we performed all calculations reported in Section \ref{sec:results} on 2 $\times$ 2 $\times$ 2 q-point meshes (corresponding to a total of $N_q$=8 q-points sampled in the irreducible wedge), using a total of $N_{\text{bands}}$=2880 (1440) energy bands for double (simple) perovskites, and a plane wave energy cutoff $\varepsilon_{\text{cutoff}}$ of 150\,eV for $\epsilon_{\bm G, \bm G'}(\bm q)$. Here, we explore the numerical errors associated with these choices for $N_q$, $N_{\text{bands}}$, and $\varepsilon_{\text{cutoff}}$.

\begin{figure}
 \includegraphics[width=\columnwidth]{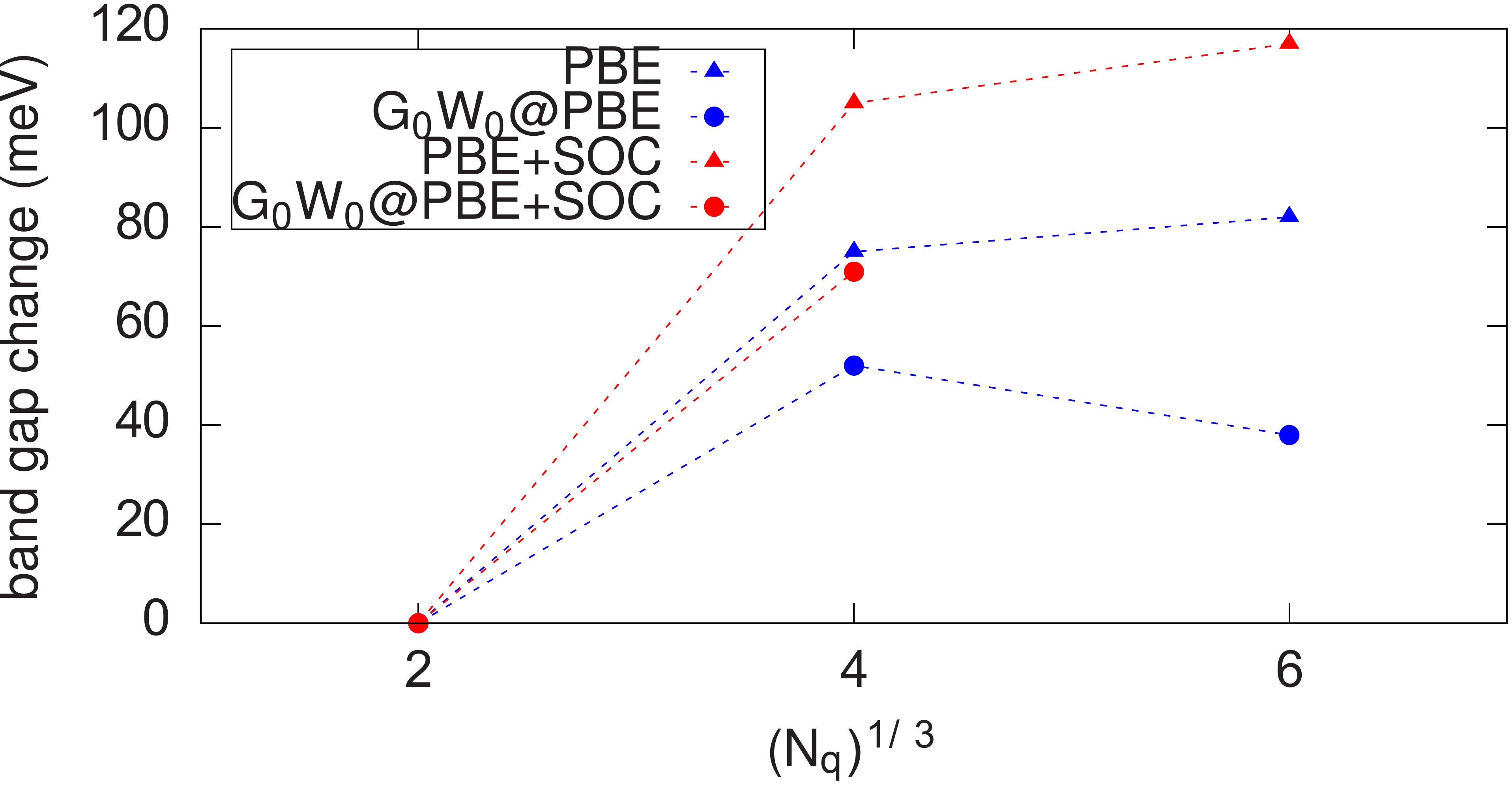}
 \caption{Change of band gap of MAPbI$_3$ as compared to the band gap calculated using an $N_q$=2 $\times$ 2 $\times$ 2 mesh, where $N_q$ is the total number of q-points sampled in the irreducible wedge, used in the construction of the dielectric function and self-energy. Dashed lines are a guide to the eye.}
  \label{fig:kpoints}
\end{figure}

\textit{Brillouin zone sampling.} Fig.~\ref{fig:kpoints} shows the change of the PBE and G$_0$W$_0$@PBE band gaps (with and without SOC) of MAPbI$_3$ for increasingly dense q-point meshes used in the construction of the dielectric function and self-energy. Upon increasing the q-point mesh from 2 $\times$ 2 $\times$ 2 to 4 $\times$ 4 $\times$ 4, the band gap changes by $\sim$50\,meV for G$_0$W$_0$@PBE and $\sim$70\,meV for G$_0$W$_0$@PBE+SOC. Using a 6 $\times$ 6 $\times$ 6 mesh leads to additional changes of $\sim$10\,meV. Additionally, we calculate the PBE band gap on an 8 $\times$ 8 $\times$ 8 mesh and extrapolate the gap linearly as a function of $1/N_q$. The extrapolated PBE band gap of MAPbI$_3$ is 1.40\,eV, compared to 1.31\,eV using the 2 $\times$ 2 $\times$ 2 grid. Based on this and the results shown in Fig.~\ref{fig:kpoints}, we estimate the error due to our finite sampling of the Brillouin zone to be below 100\,meV for MAPbI$_3$. In the case of (MA)$_2$BiTlBr$_6$, the difference between the G$_0$W$_0$@PBE band gap calculated on a 2 $\times$ 2 $\times$ 2 and a 4 $\times$ 4 $\times$ 4 mesh is 95\,meV, suggesting a similar finite q-point grid error as for MAPbI$_3$.

\begin{figure}
 \includegraphics[width=\columnwidth]{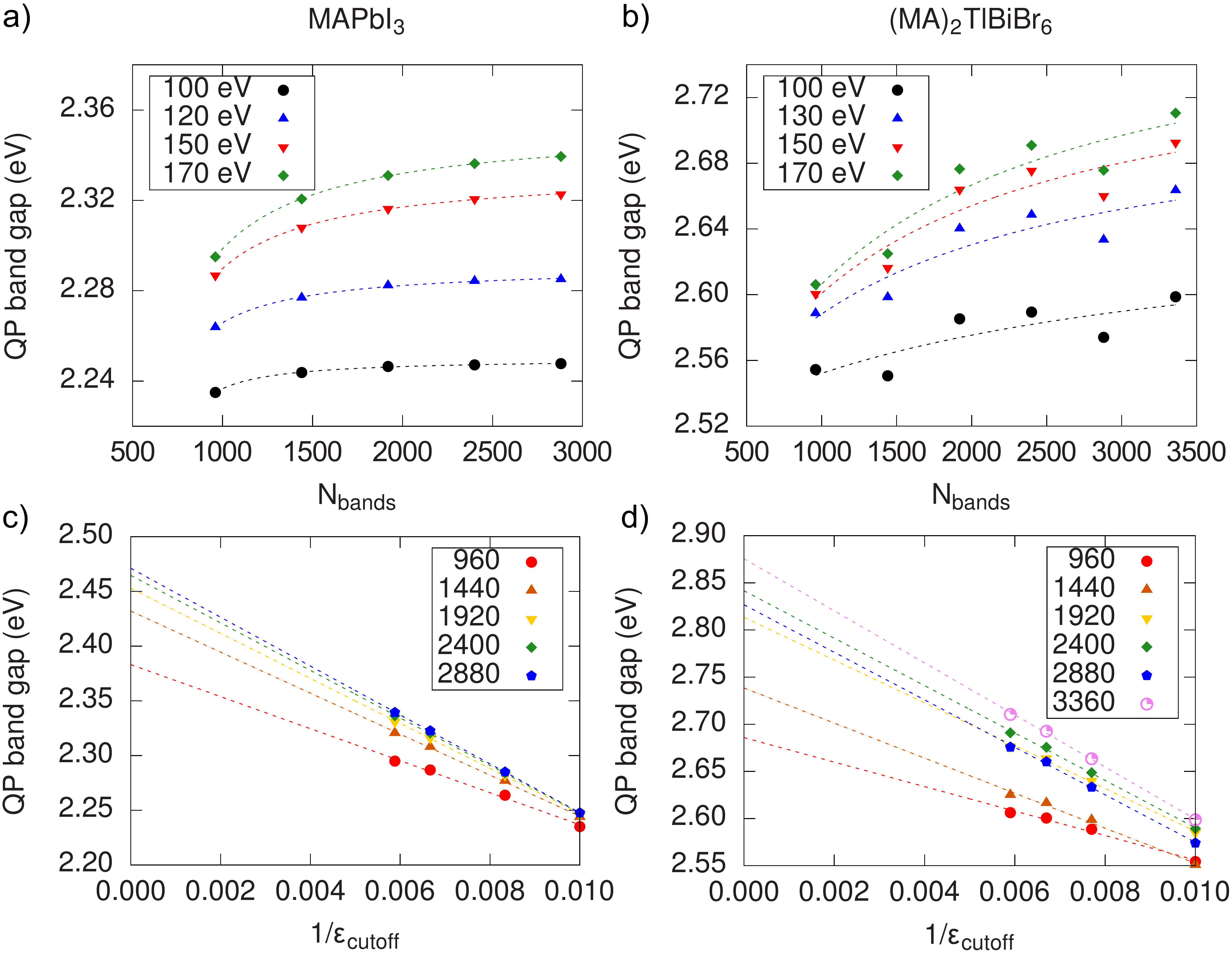}
 \caption{a) and b) G$_0$W$_0$@PBE band gap of MAPbI$_3$ and (MA)$_2$BiTlBr$_6$ without SOC as a function of the total number of bands and $\varepsilon_{\text{cutoff}}$. Dashed lines correspond to fits using Eq. \ref{ref:hyperbolical}, c) and d) G$_0$W$_0$@PBE band gap of MAPbI$_3$ and (MA)$_2$BiTlBr$_6$ as a function of $1/\varepsilon_{\text{cutoff}}$ and for different $N_{\text{bands}}$. Dashed lines correspond to linear fits of data points.}
  \label{fig:conv}
\end{figure}

\textit{Number of empty bands.} Our tests of the convergence of the G$_0$W$_0$@PBE band gap with respect to the number of empty bands used in the construction of the dielectric function and the self-energy neglect SOC and use PAW2 potentials (see Table \ref{tab:pseudos}) for both MAPbI$_3$ (Fig.~\ref{fig:conv}a) and (MA)$_2$BiTlBr$_6$ (Fig.~\ref{fig:conv}b). Following earlier work \cite{Friedrich2011}, we find that the function
\begin{equation}
\label{ref:hyperbolical}
    f(N_{\text{bands}}) = \frac{a}{N_{\text{bands}} - N_0}+b,
\end{equation}
where $a$, $b$ and $N_0$ are fit parameters, well describes our calculations for MAPbI$_3$ (see Fig.~\ref{fig:conv}a). We therefore estimate the error due to using a finite number of empty bands by identifying the asymptote $b$ with the band gap extrapolated to infinite $N_{\text{bands}}$. For an $\varepsilon_{\text{cutoff}}$ of 150\,eV, the difference between the extrapolated and the $N_{\text{bands}}$=1440 G$_0$W$_0$@PBE band gap is 26\,meV. For (MA)$_2$BiTlBr$_6$ (Fig.~\ref{fig:conv}b), using the same procedure, we find that the error due to using $N_{\text{bands}}$=2880 is 97\,meV.

\textit{Dielectric matrix cutoff.} The convergence of the G$_0$W$_0$@PBE band gap with respect to $1/\varepsilon_{\text{cutoff}}$, where $\varepsilon_{\text{cutoff}}$ is the plane wave cutoff used to describe $\epsilon_{\bm G, \bm G'}(\bm q)$, is shown in Fig.~\ref{fig:conv}c and \ref{fig:conv}d for MAPbI$_3$ and (MA)$_2$BiTlBr$_6$, respectively. We linearly extrapolate the QP band gap to infinite energy cutoffs. For MAPbI$_3$ ((MA)$_2$BiTlBr$_6$), the extrapolated G$_0$W$_0$@PBE band gap for 1440 (2880) bands is 2.43\,eV (2.83\,eV). Our use of $\varepsilon_{\text{cutoff}}$=150\,eV is therefore expected to underestimate the band gap of 0.12\,eV for MAPbI$_3$ and 0.17\,eV for (MA)$_2$BiTlBr$_6$, respectively.

Further convergence parameters are the cutoff energy for the plane wave expansion of the wavefunctions and the number of grid points used for the frequency integration of the screened Coulomb interaction, for which we use values of 500\,eV and 100 grid points, respectively. Increasing these parameters changes QP band gaps by less than 40\,meV. Given the above, we estimate that our QP band gaps are underestimated by $\sim$0.2\,eV for simple and up to $\sim$0.4\,eV for double perovskites due to using finite $N_q$, $N_{\text{bands}}$, and $\varepsilon_{\text{cutoff}}$.

\subsection{Effect of crystal structure}
\label{subsec:structure}
The choice of crystal structure can account for differences of several 100\,meV in the predicted band gap of hybrid halide perovskites. We demonstrate this by comparing the PBE and G$_0$W$_0$@PBE band gaps of MAPbBr$_3$ using the experimental and a geometry-optimized structure from Ref.~\citenum{Bokdam2016}, shown in Fig.~\ref{fig:unitcell}c and \ref{fig:unitcell}d. In both structures, we replace MA by Cs, but keep the unit cell volume and atomic positions of the PbBr$_6$ octahedra unchanged. For the optimized structure from Ref.~\citenum{Bokdam2016} we find band gaps of 1.03\,eV (PBE) and 2.00\,eV (G$_0$W$_0$@PBE), $\sim$0.5\,eV larger than with the experimental structure (PBE and G$_0$W$_0$@PBE band gaps in Table \ref{tab:gaps}). These differences can be attributed to spurious distortions of the PbBr$_6$ octrahedra in the geometry-optimized structure, e.g., off-center displacements of Pb and octahedral tilts, which break the symmetry of the crystal and tend to open the band gap \cite{Zheng2015,Leppert2016}. Due to strong coupling of the MA molecule's dipole moment and the PbBr$_6$ octahedra, such distortions can occur during geometry optimization and sensitively depend on the chosen orientation of the molecules and the volume of the unit cell. The effect of local symmetry-breaking and the complex dynamical structure of hybrid halide perovskites are debated in the literature. Recent experimental work\cite{Frohna2018a} based on second harmonic generation rotational anisotropy has confirmed earlier X-ray diffraction measurements \cite{Kawamura2002, Weller2015a} that had assigned a centrosymmetric \textit{I4/mcm} space group to MAPbI$_3$ at room temperature, and found no evidence of local symmetry breaking. Furthermore, bimolecular charge carrier recombination in MAPbI$_3$, which had previously been linked to static and dynamical symmetry breaking in halide perovskites \cite{Motta2015b,Zheng2015, Etienne2016a}, has been shown to be due to direct, fully radiative band-to-band transitions, also suggesting a centrosymmetric space group \cite{Davies2018}. However, polar fluctuations have been observed in both hybrid organic-inorganic and all-inorganic halide perovskites \cite{Yaffe2017}, and the effect of dynamical fluctuations on the band gaps of several halide perovskites has been discussed in Refs.\citenum{Wiktor2017a} and \citenum{McKechnie2018}, and shown to be as high as $\sim$0.3\,eV for MAPbI$_3$. While further investigations of the complex dynamical structure of halide perovskites and its influence on optoelectronic properties are warranted, our work is focused on the predictive power of GW calculations, and for reasons of consistency, all band gaps reported in the following are obtained using experimental, on-average centrosymmetric, structures.

\section{Results and discussion}
\label{sec:results}
The PBE band gaps of all compounds are reported in Table \ref{tab:gaps}. Cs$_2$BiAgBr$_6$ has an indirect fundamental band gap with the VBM at X=($\frac{1}{2}$,0,$\frac{1}{2}$) and the CBM at L=($\frac{1}{2}$,$\frac{1}{2}$,$\frac{1}{2}$). Cs$_2$Au$_2$I$_6$ has a direct band gap along the high-symmetry line from $\Sigma$=(0.36,0.64,-0.36) to $\Gamma$, only 0.02\,eV lower in energy than the direct gap at N=(0,$\frac{1}{2}$,0); we use the direct gap in what follows. All other systems studied here have direct band gaps at R=($\frac{1}{2}$,$\frac{1}{2}$,$\frac{1}{2}$) (simple perovskites) or $\Gamma$ (double perovskites). PBE severely underestimates the experimental band gaps by more than 1\,eV for all compounds, and by up to almost 2\,eV for MAPbBr$_3$ and Cs$_2$TlAgCl$_6$. For Cs$_2$TlAgBr$_6$, PBE incorrectly predicts a "negative band gap", i.e., a crossing of bands at the Fermi level \citep{Slavney2018a}. We note that these results further demonstrate that trends in PBE band gaps are not always reliable for halide perovskites: For example, the PBE band gap of (MA)$_2$BiTlBr$_6$ is $\sim$0.5\,eV smaller than that of Cs$_2$BiAgBr$_6$, whereas the experimental band gaps of these compounds are almost the same.

\begin{sidewaystable*}
\caption{gKS and QP band gaps (in eV) compared to experimental (optical) band gaps. GW$_0$ refers to eigenvalue self-consistent calculations as described in the text. The negative PBE band gap is calculated as the energy difference between the minimum of the band identified as the CBM and the maximum of the band identified as the VBM in a separate HSE06 band structure calculation\cite{Slavney2018a}. G$_0$W$_0$@HSE06 and GW$_0$@HSE06 band gaps for Cs$_2$InAgCl$_6$, Cs$_2$SnBr$_6$, and Cs$_2$Au$_2$I$_6$ were not calculated, because their GW$_0$@PBE band gaps are in very good agreement with experiment.}
\begin{center}
\begin{tabular}{ccccccccccc}
\hline \hline
system & PBE & G$_0$W$_0$@PBE & GW$_0$@PBE & HSE & G$_0$W$_0$@HSE & GW$_0$@HSE & PBE0 & G$_0$W$_0$@PBE0 & GW$_0$@PBE0 & exp \\ \hline 
MAPbI$_3$  & 0.21 & 0.94 & 1.08 & 0.82 & 1.26 & 1.33 & 1.53 & 1.64 & 1.65 & 1.52\cite{Stoumpos2013a}, 1.69\cite{Quarti2016} 
\\ 
MAPbBr$_3$  & 0.55 & 1.49 & 1.68 & 1.30 & 2.01 & 2.10 & 2.03 & 2.37 & 2.42 & 2.35\cite{Kitazawa2002},
2.30\cite{Ryu2014a},
2.39\cite{Yang2015b}
\\ 
CsSnBr$_3$  & 0.06 & --- & --- & 0.63 & 1.02 & 1.14 & 1.36 & 1.37 & 1.34 & 1.75\cite{Gupta2016a}
\\ \hline 
(MA)$_2$BiTlBr$_6$ & 0.60 & 1.40 & 1.55 & 1.00 & 1.78 & 2.00 & ---& ---& ---& 2.16\cite{Slavney2017,Deng2016c}
\\ 
Cs$_2$TlAgBr$_6$ & -0.66 & --- & --- & 0.20 & 0.63 & 0.82 & ---& ---& ---& 0.95\cite{Slavney2018a} 
\\ 
Cs$_2$TlAgCl$_6$ & 0.00 & --- & --- & 1.09 & 1.87 & 2.17 & ---& ---& ---& 1.96\cite{Slavney2018a}
\\ 
Cs$_2$BiAgBr$_6$ & 1.09 & 2.01 & 2.22 & 1.95 & 2.59 & 2.82 & ---& ---& ---& 1.95\cite{Slavney2016a},
2.19\cite{McClure2016}
\\ 
Cs$_2$InAgCl$_6$ & 1.16 & 2.79 & 3.12 & 2.61 & --- & --- & ---& ---& ---& 3.23\cite{Zhou2017}, 3.30\cite{Volonakis2017}, 3.53\cite{Tran2017} 
\\ 
Cs$_2$SnBr$_6$ & 1.10 & 2.96 & 3.16 & 2.14 & --- & --- & --- & --- & --- & 2.70\cite{Kaltzoglou2016}, 2.85\cite{Lee2017a}, 3.00\cite{Dalpian2017}
\\ 
Cs$_2$Au$_2$I$_6$ & 0.70 & 1.28 & 1.40 & 1.16 & --- & --- & --- & --- & --- & 1.31\cite{Debbichi2018}
\\ \hline \hline
\end{tabular}
\end{center}
\label{tab:gaps}
\end{sidewaystable*}
In Table \ref{tab:gaps}, we present our calculated G$_0$W$_0$@PBE band gaps for all compounds, except for the zero/"negative" band gap compounds Cs$_2$TlAgX$_6$ (X=Br, Cl) and CsSnBr$_3$. In line with previous studies, we find that G$_0$W$_0$@PBE underestimates the measured band gap of MAPbI$_3$ by $\sim$0.7\,eV \citep{Brivio2014}. For MAPbBr$_3$ the underestimation with respect to experiment is even more severe. In contrast, we find a G$_0$W$_0$@PBE band gap of 2.01\,eV for Cs$_2$BiAgBr$_6$, in excellent agreement with experiment. We note that the experimental uncertainty for band gaps of halide perovskites is significant, up to $\sim$0.3\,eV for some of the systems studied here (Table \ref{tab:gaps}).

Our results are in line with previous reports for MAPbI$_3$ \cite{Brivio2014,Bokdam2016}, where the use of self-consistency in the GW calculations was concluded to be crucial to obtain band gaps in agreement with experiment. However, while some prior reports used a QP self-consistent scheme (QSGW) in which both eigenvalues and eigenfunctions are updated self-consistently\cite{Huang2013,Brivio2014,Wiktor2017a}, others have argued that partial self-consistency in the eigenvalues (GW$_0$) is sufficient for MAPbX$_3$ (X=I, Br, Cl) \cite{Bokdam2016}. Here, we test the effect of eigenvalue self-consistency on the band gap for a broader class of halide perovskites than has been considered previously. For most compounds, our GW$_0$@PBE band gaps are converged to within less than 0.05\,eV after four iterations. We find that eigenvalue self-consistency in G increases the band gap by a maximum of $\sim$0.2\,eV for MAPbI$_3$, MAPbBr$_3$, CsSnBr$_3$, (MA)$_2$BiTlBr$_6$, Cs$_2$BiAgBr$_6$, Cs$_2$SnBr$_6$, and Cs$_2$Au$_2$I$_6$, and by 0.3\,eV for Cs$_2$InAgCl$_6$. For the former four compounds, our GW$_0$ calculations continue to significantly underestimate the band gap relative to experiment. 

Our calculations demonstrate that GW$_0$@PBE with eigenvalue self-consistency is not a general approach for predicting reliable band gaps across all classes of halide perovskites. As a consequence, the assumption $\psi_{nk}^{\text{QP}} \approx \psi_{nk}^{\text{PBE}}$ is likely not justified. We therefore turn to the hybrid functional HSE06 \cite{Krukau2006} to obtain a better approximation to $\psi_{nk}^{\text{QP}}$ and a better starting point for our G$_0$W$_0$ calculations. Our calculated HSE06, G$_0$W$_0$@HSE06 and GW$_0$@HSE06 band gaps are reported in Table \ref{tab:gaps}. As expected, due to its inclusion of a fraction of non-local exact exchange, HSE06 opens up the band gap considerably for all compounds. For MAPbBr$_3$, CsSnBr$_3$ and (MA)$_2$BiTlBr$_6$, DFT-HSE06 continues to underestimate the experimental band gap by $\sim$1\,eV. GW$_0$@HSE06 leads to band gaps within $\sim$0.2\,eV of the experimental gap for almost all systems. However, it overshoots the experimentally measured gap of Cs$_2$BiAgBr$_6$ by $\sim$0.8\,eV, illustrating that HSE06, is not a one-size-fits-all solution either. 

For the simple perovskites MAPbX$_3$ (X=I, Br) and CsSnBr$_3$, we also calculate the PBE0, G$_0$W$_0$@PBE0 and GW$_0$@PBE0 band gaps (Table \ref{tab:gaps}). For all three systems, PBE0 underestimates the experimental gap by less than $\sim$0.4\,eV. We find that the GW$_0$@PBE0 band gaps of MAPbX$_3$ (X=I, Br) are in excellent agreement with experiment; the GW$_0$@PBE0 band gap of CsSnBr$_3$ underestimates the measured gap by 0.4\,eV.

\begin{figure}[t]
 \includegraphics[width=\columnwidth]{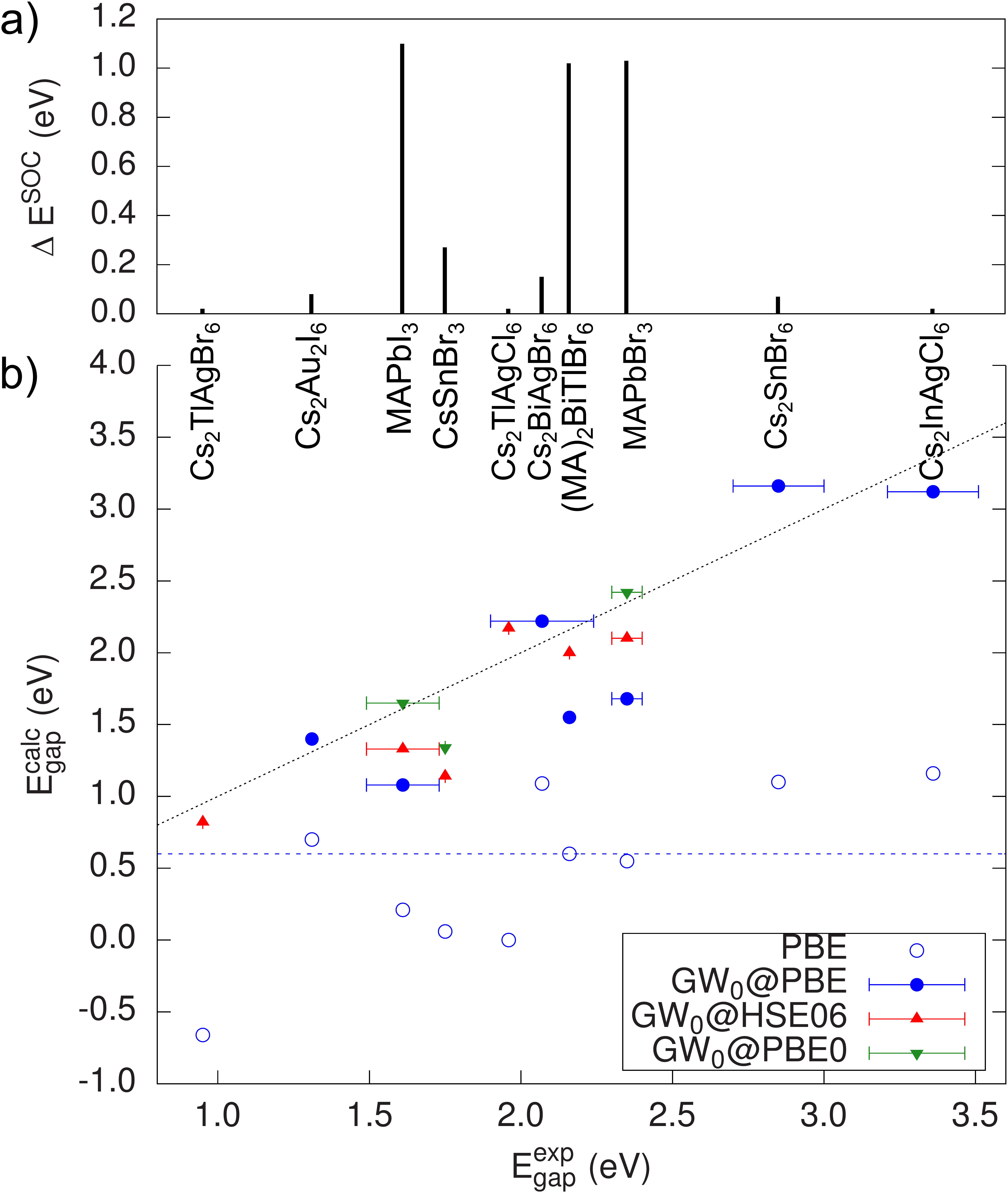}
 \caption{a) Difference between DFT-PBE band gaps calculated with and without SOC. b) Calculated fundamental versus experimental band gaps. We plot the average band gap with the standard deviation as an error bar, when a range of values is reported in the literature.}
  \label{fig:bandgaps}
\end{figure}

The starting point dependence in the systems studied here is indicative of overscreening, a phenomenon previously pointed out in the context of G$_0$W$_0$ calculations on organic molecules \citep{Blase2011}; overscreening is particularly significant for systems with very small or even vanishing PBE band gaps. For the systems studied here, small PBE band gaps can be a result of large SOC, where here we assess the impact of SOC qualitatively as $\Delta E^{\text{SOC}} = E_{\text{gap}}^{\text{w/o SOC}} - E_{\text{gap}}^{\text{with SOC}}$, shown in Fig.~\ref{fig:bandgaps}a. This is the case for MAPbI$_3$, (MA)$_2$BiTlBr$_6$, and MAPbBr$_3$. However, small band gaps can also occur in systems with small SOC, and originate from a large dispersion of the CBM and a small energy difference between the atomic orbitals from which the CBM and VBM are derived, as discussed in Ref.~\citenum{Slavney2018a}. This is the case for Cs$_2$TlAgX$_6$ (X=Br, Cl) and CsSnBr$_3$. All of these systems have PBE band gaps smaller than $\sim$0.6\,eV, and their GW$_0$@PBE band gaps show large deviations from experiment (Fig.~\ref{fig:bandgaps}b). 

Based on our results, we assign the halide perovskites we have studied here to two groups. Group I contains well-known simple perovskites such as MAPbI$_3$ and the closely related MAPbBr$_3$ with Pm$\bar{3}$m symmetry, as well as the double perovskites (MA)$_2$BiTlBr$_6$ and Cs$_2$TlAgX$_6$ (X=Br,Cl). Other perovskites either containing heavy elements or structural geometries and orbital hybridization that favor highly dispersive band edges and comparably low band gaps will likely also belong to this group. For these materials, gKS eigensystems based on semilocal exchange-correlation functionals such as PBE lead to very small or even vanishing gKS band gaps. Hybrid functionals like HSE06, in the absence of GW corrections, improve the situation somewhat, but can still underestimate the experimental band gaps by up to $\sim$1\,eV. Likewise, one-shot G$_0$W$_0$@PBE corrections do not significantly increase the band gap relative to experiment, and neither does self-consistency in the QP eigenvalues. Band gaps computed with GW$_0$@HSE06 lead to a dramatic improvement for most compounds in this group, as shown in Fig.~\ref{fig:bandgaps}b. However, even HSE06 is not sufficient for all Group I materials, most notably MAPbI$_3$, where GW$_0$@HSE06 does not reconcile calculated and experimental band gaps. The crucial role of balanced, system-dependent contributions of semilocal and nonlocal exchange and correlation is further demonstrated by our GW$_0$@PBE0 calculations for MAPbX$_3$ (X=I, Br) and CsSnBr$_3$ which lead to band gaps in very good agreement with experiment for the former while still underestimating the measured gap of the latter (see Fig.~\ref{fig:bandgaps}b).

 A promising approach could be to generate gKS starting points for G$_0$W$_0$ or GW$_0$ self-consistently for each system, for example by using optimally-tuned range separated hybrids (OTRSHs), in which a tunable parameter determines the range at which long-range exact exchange and short-range semilocal exchange dominate, as discussed in Ref.~\citenum{Yanai2004}. Additionally, enforcing the correct asymptotic limit of the potential of $1/(\epsilon_{\text{mac}}^{\infty} r)$ has been shown to yield excellent band gaps in molecular crystals \cite{Refaely-Abramson2013}, where $\epsilon_{\text{mac}}^{\infty}$ is the electronic contribution to the average macroscopic dielectric constant.  Alternatively, in a global hybrid functional approach one could explore setting the fractional Fock exchange $\alpha \approx 1/\epsilon_{\text{mac}}^{\infty}$, as proposed in Ref.~\citenum{Skone2014}, and proceed with this gKS starting point.

The all-inorganic double perovskites Cs$_2$BiAgBr$_6$, Cs$_2$InAgCl$_6$, Cs$_2$Au$_2$I$_6$, and Cs$_2$SnBr$_6$ -- and likely other perovskites with weaker spin-orbit interactions, and orbital hybridization or octahedral distortions favoring larger band gaps -- belong to Group II. Here, although PBE does underestimate the band gap, it represents an adequate starting point for G$_0$W$_0$ or GW$_0$ calculations for this group of systems.

Our findings suggest that one-shot or eigenvalue self-consistent GW calculations from hybrid functional starting points can improve band gaps of some (although not all) halide perovskites relative to experiment. Our work has important consequences for comparative studies of doped or alloyed systems, investigating, e.g., the band gap evolution as a function of dopant or alloy concentration. Reliable QP energies are also a necessary prerequisite for the quantitative prediction of defect levels and excited state properties, such as optical band gaps and exciton binding energies. As we have shown here, to obtain quantitative gaps with GW for halide perovskites, hybrid functional-based starting points or full self-consistency can be necessary.

\begin{acknowledgments}
This work was supported by the National Science Foundation under Grant No. DMR-1708892. Portions of this work were also supported by the Molecular Foundry through the U.S. Department of Energy, Office of Basic Energy Sciences under Contract No. DE-AC02-05CH11231. L.L. acknowledges partial support by the Feodor-Lynen program of the Alexander-von-Humboldt foundation, by the Bavarian State Ministry of Science and the Arts for the Collaborative Research Network "Solar Technologies go Hybrid (SolTech)", the Elite Network Bavaria, and the German Research Foundation (DFG) through SFB840.
\end{acknowledgments}

\end{document}